\documentclass[prd,showpacs,nofootinbib,tightenlines,preprintnumbers,superscriptaddress,twocolumn]{revtex4}
\usepackage{epsfig,graphics}
\usepackage{amssymb,amsmath,amsthm,graphicx,psfrag}

\newcommand{\beqn}{\begin{eqnarray}}
\newcommand{\eeqn}{\end{eqnarray}}
\newcommand{\be}{\begin{equation}}
\newcommand{\ee}{\end{equation}}
\newcommand{\eq}[1]{(\ref{#1})}

\begin{document}

\preprint{HU-EP-04/66}
\preprint{ITEP-LAT/2004-18}

\title{The monopole content of topological clusters : \\ 
have KvB calorons been found?}

\author{E.-M. Ilgenfritz}
\affiliation{Institut f\"ur Physik, Humboldt-Universit\"at zu Berlin,
Newtonstr. 15, D-12489 Berlin, Germany}
\author{B. V. Martemyanov}
\affiliation{Institute for Theoretical and Experimental Physics,
B. Cheremushkinskaya 25, Moscow 117259, Russia}
\author{M. M\"uller-Preussker}
\affiliation{Institut f\"ur Physik, Humboldt-Universit\"at zu Berlin,
Newtonstr. 15, D-12489 Berlin, Germany}
\author{A. I. Veselov}
\affiliation{Institute for Theoretical and Experimental Physics,
B. Cheremushkinskaya 25, Moscow 117259, Russia}


\begin{abstract}
Using smearing of equilibrium lattice fields generated at finite 
temperature in the confined phase of $SU(2)$ lattice gauge theory, 
we have investigated the emerging topological objects (clusters of 
topological charge).
Analysing their monopole content according to the Polyakov gauge and
the maximally Abelian gauge, we characterize part of them to correspond
to nonstatic calorons or static dyons in the context of Kraan-van Baal 
caloron solutions with non-trivial holonomy.
The behaviour of the Polyakov loop inside these clusters and the 
(model-dependent) topological charges of these objects support this 
interpretation. 
\end{abstract}


\maketitle


\section{Introduction}
\label{sec:introduction}
The space-time distribution of topological charge in the Euclidean
vacuum, as it can be made visible in lattice gauge theory (LGT), 
has continued to be an interesting topic, being the source of chiral 
symmetry breaking and a manifestation of the $U_A(1)$ anomaly. 
In this context, the relation to the confining property of the vacuum 
has been mostly left out of consideration. 
The guiding line of this activity was the instanton liquid model~\cite{ILM}, 
originally motivated by the semiclassical approximation, which then has turned 
into a successful and sufficiently rich model of the gluonic background 
of hadron physics.~\footnote{Recently, however, also dissident views have been 
developed~\cite{Horvath}.} The inability of the instanton model 
to describe confinement was not considered as an essential disadvantage.
The semiclassical background of the model itself has suggested to employ 
(limited) cooling as the method of choice to detect the background fields 
which actually revealed themselves as consisting of lumps of action and 
topological charge.

The main problem was to find the density and size distribution of these 
topological objects. 
Such studies have relied on very subjective tools of smoothing the 
UV gauge field fluctuations (cooling~\cite{ILMPSS,Teper,PV,dFGPS}, cycles 
of blocking and inverse blocking~\cite{smoothing-1,smoothing-2,smoothing-3}, 
four-dimensional smearing~\cite{smearing}). Whereas the existence of 
``hot spots'' by themselves 
(very localized regions of strong field strength, where the field turns out 
to be approximately selfdual or antiselfdual) was undoubtedly an outstanding 
feature of smoothing, the number of these lumps and (less strongly) their 
sizes were depending on details and prejudices.

Recent developments lead to the impression that this might not be the final 
word:

\begin{itemize}
\item The notion of calorons (instantons at finite temperature) has been 
extended to more complicated solutions (KvB calorons) in a background of 
non-trivial holonomy~\cite{KvB-1,KvB-2,LL}. 
An important part of the moduli space corresponds to calorons dissociated 
into constituents (BPS monopoles or dyons). The instability with respect to 
dissociation has been discussed in the context of the transition from 
deconfinement to confinement~\cite{Brower,Diakonov1,Diakonov2}.
Moreover, during the last years it has been found that on asymmetric lattices, 
starting from lattice ensembles in the confinement phase, cooling leads 
to configurations which resemble single KvB calorons or a gas of calorons 
and caloron constituents~\cite{IMMPSV,IMMPV4,FBIMPvB}. 

\item There is a strong desire to built models for fully non-Abelian gauge 
fields (opposed to models relying on an Abelian or center 
projection~\cite{Engelhardt}) which would be able to describe confinement 
together with chiral symmetry breaking based on topologically charged 
objects (in order to realize the $U_A(1)$ anomaly). This has recently led 
to a reconsideration of the instanton liquid model, in this case starting 
from gauge field configurations with topological charge spread out over
{\em large} portions of space, giving rise to colour correlations over
large distances~\cite{Negele}.
\end{itemize}

A systematic consideration of ensembles of KvB calorons in the respective
holonomy background would also be motivated by this objective and is hoped 
finally to provide a semiclassically motivated model~\cite{Diakonov1} working 
in the neighbourhood of the confinement/deconfinement phase transition.

In our previous papers~\cite{IMMPSV,IMMPV4,FBIMPvB}
we mainly concentrated on almost classical lattice 
configurations, calorons and constituent dyons and antidyons,
obtained by cooling. Of course, 
the non-trivial holonomy necessary to find the most interesting new types 
of classical solutions has changed during cooling compared with the holonomy 
of the corresponding Monte Carlo configuration, but sufficiently manifold
configurations (with various topological charges $Q$) have been found which 
should be typical for calorons corresponding to a confining background.

In this paper we want to analyse lattice configurations closer to the 
equilibrium (Monte Carlo) ensemble by gluonic observables.  This requires 
to replace cooling, usually minimizing the action down to the level of 
classical configurations, by 
four-dimensional smearing~\cite{smearing}. At this level of 
smoothing the distribution of topological charge becomes visible in the 
form of more general clusters of topological charge. It should be said
that, even without any smoothing, the low-lying eigenmodes of  sufficiently
chirally improved fermions are a valuable tool to decipher the topological 
structure of individual gauge field configurations. Restricting the attention 
to $SU(3)$ configurations with $Q=\pm 1$, the corresponding single fermionic 
zero mode (with boundary conditions manipulated at will) has been 
demonstrated~\cite{Gattringer} to be an ideal tool for localizing 
(possibly dyonic) caloron constituents. In a joint attempt to confirm this 
interpretation~\cite{Lattice2003} we have 
applied smearing to a set of lattice configurations 
from this study and have used the gluonic topological density in order to 
prove that the zero mode jumps indeed between clusters of topological charge. 
The exact dyonic nature of the clusters, however, remained inconclusive.

In this paper we apply smearing to an ensemble of finite temperature
confining $SU(2)$ configurations. In the line of our previous 
studies~\cite{IMMPSV} and similar observations for $SU(3)$ 
calorons~\cite{Peschka} we will classify the clusters of topological 
charge with respect to the content of Abelian monopoles, both in the Polyakov 
gauge and in the maximally Abelian gauge. 
We give evidence that this classification can be understood as identification 
of some clusters as dissociated (charge $|Q_{cluster}| \approx 1/2$) and 
undissociated (charge $|Q_{cluster}| \approx 1$) ones with an internal 
structure of the Polyakov loop resembling the corresponding limiting cases 
of KvB caloron solutions.

We recall that correlations between Abelian monopoles and topological density
have been studied already in the past, both without~\cite{correl1} and 
with smoothing~\cite{correl2,correl3,Gubarev}. Here we go a step further and 
use the location and number of Abelian monopoles in order to see the 
correlation with the Polyakov loop variable inside topological clusters.

The paper is organized as follows. In Section \ref{sec:smearing} we describe
the sample of lattice configurations, smearing and basic local features of
the Polyakov loop. In Section \ref{sec:cluster} we describe the cluster
analysis which allows us to extract specific statistical properties of
clusters interpreted as caloron constituents and undissociated calorons.
In Section \ref{sec:conclusion} we draw our conclusions.

\section{Description of the method}
\label{sec:smearing}
By the Monte Carlo method we have generated 500 configurations on
a $20^3\times 6$ lattice at $\beta = 2.3$. This sample characterizes finite 
temperature still in the confined phase. The ensemble further underwent 
smoothing by four-dimensional smearing~\cite{smearing}. The fixed smearing 
parameter was $\alpha = 0.45$, whereas various numbers $N$ of iterations 
have been investigated. One iteration of smearing corresponds to replacing each 
$j$-level smeared link of the lattice by the $(j+1)$-level smeared link which 
is obtained as the normalized superposition of the $j$-level link 
(with weight $1-\alpha$) and the surrounding 6 staples of $j$-level links 
(each with weight $\alpha/6$). 	
Typical numbers of iterations were $N=50$ and $N=100$ smearing steps upon 
which the action of an initially thermalized configuration became reduced by 
a factor $1/150$ and $1/300$ to values $S \approx (30 \ldots 50) S_{inst}$
and $S \approx (15 \ldots 25) S_{inst}$, respectively. $S_{inst}$ is the
action of an instanton or caloron with one unit of topological charge.
In this stadium of smearing the initially uniformly noisy configurations have 
developed into configurations containing clusters of topological charge. 
Near the maxima of action density the field strength is approximately 
(anti-)selfdual. For the smeared configurations we have recorded the profiles 
of the action density, the topological charge density, the Polyakov loop, 
and have located the trajectories of Abelian monopoles~\cite{dGT}
obtained after Abelian projection, either in Polyakov gauge (PG) or 
in the maximal Abelian gauge (MAG)~\cite{tHooft,Schierholz}. 
The Polyakov gauge is characterized by diagonalizing the Polyakov loops 
(and, consequently, by diagonal time-like links). 

Employing these observables we have searched for signatures of KvB solutions
among the clusters of topological charge. We have mostly concentrated on those  
configurations that have maximally non-trivial holonomy, i.e. the
trace of the Polyakov loop should behave as $L(\vec{x}) \to H \simeq 0~$ 
for $~|\vec{x}-\vec{x}_i| \gg |\vec{x}_1-\vec{x}_2|$, where 
$~\vec{x}_i, i=1,2~$ denote the positions of the two centers of the $SU(2)$
KvB calorons. Moreover, we were searching for the two limiting cases of 
well-separated dyon pairs and undissociated calorons characterized as 
follows~\cite{KvB-1,KvB-2,IMMPSV}:
\begin{itemize}
\item 
Static calorons dissociate into two separate lumps, BPS monopoles or 
dyons with approximately half-integer topological charge each. 
The Polyakov loop has an equal, but opposite sign throughout each of the 
two lumps of topological charge 
and is peaking very close to the positions of local maxima of the action 
or topological charge density with
$L(\vec{x}_i) = \pm 1$. Each static dyon contains a 
(static) Abelian (anti)monopole world line at the center closing 
through the periodic boundary condition in the imaginary time direction.   
\item
Undissociated calorons are seen as connected clusters of topological charge. The
Polyakov loop changes sign within the cluster and has peaks $L \to \pm 1$
in the neighbourhood of the center of action density, i.e. it exhibits a 
dipole-like structure inside the cluster. 
On the other hand the Abelian monopole world lines close
within the cluster and not via the periodic boundary condition. The overall
magnetic charge in the part of 3-space where it is 
intersecting the cluster is expected to cancel.
The occurence of Abelian monopoles in any case is taken as an indicator for 
the local positions of the non-Abelian caloron constituents.
\end{itemize}

\begin{figure*}
\vspace*{0.7 cm}
\centering
\includegraphics[width=.31\textwidth,height=.31\textwidth]{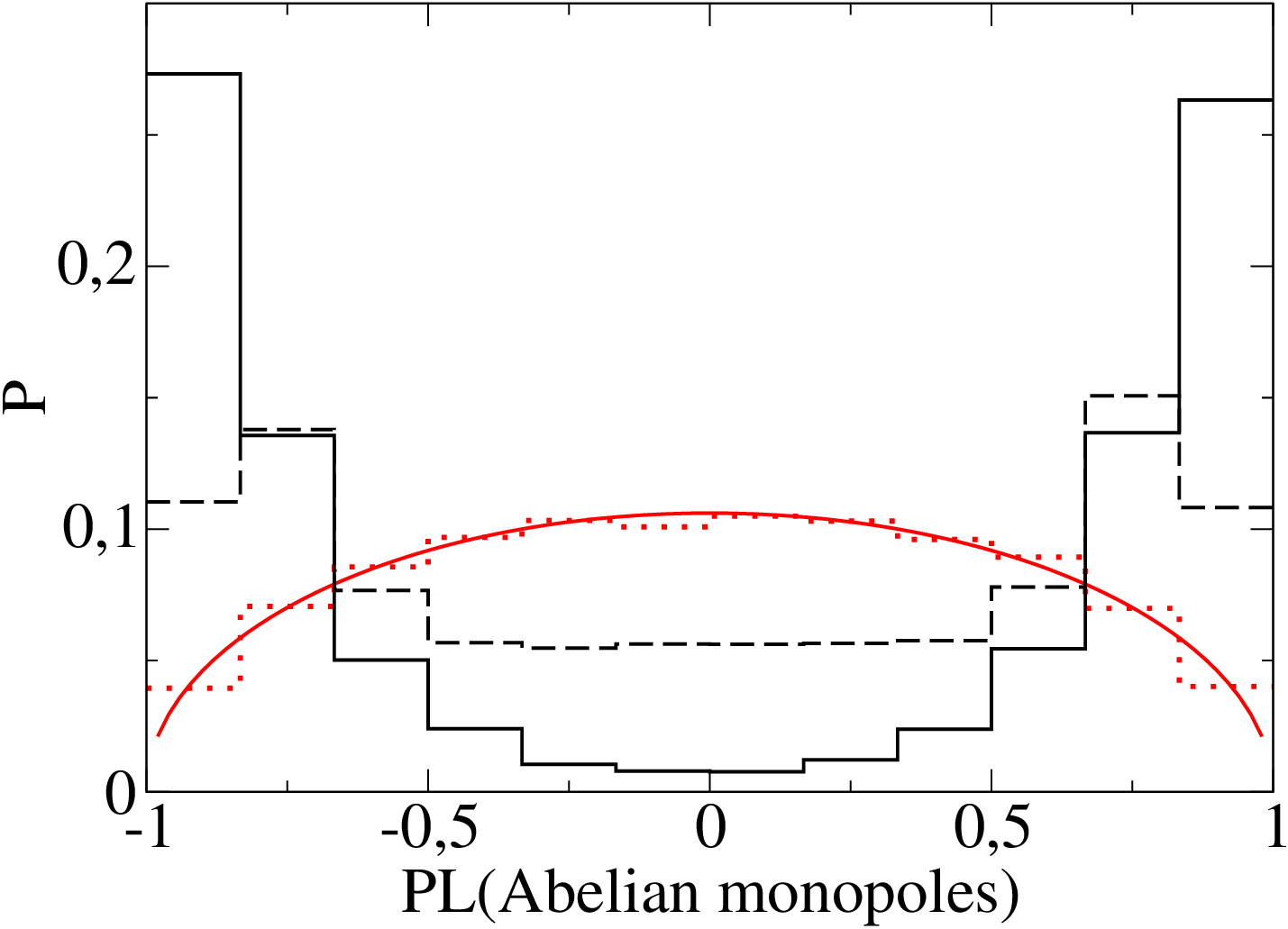}%
\hspace{.3 cm} \includegraphics[width=.31\textwidth,height=.31\textwidth]{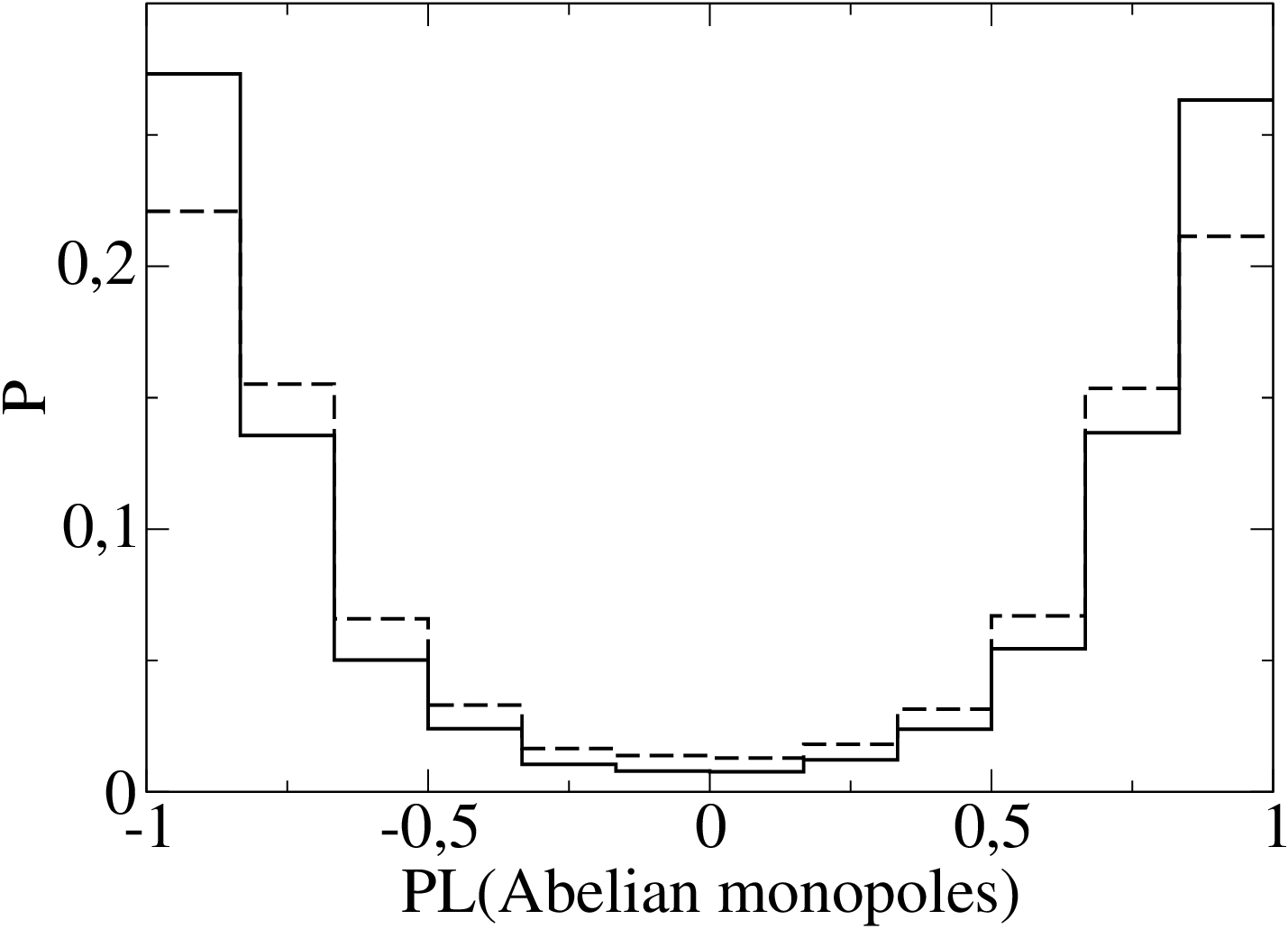}%
\hspace{.3 cm} \includegraphics[width=.31\textwidth,height=.31\textwidth]{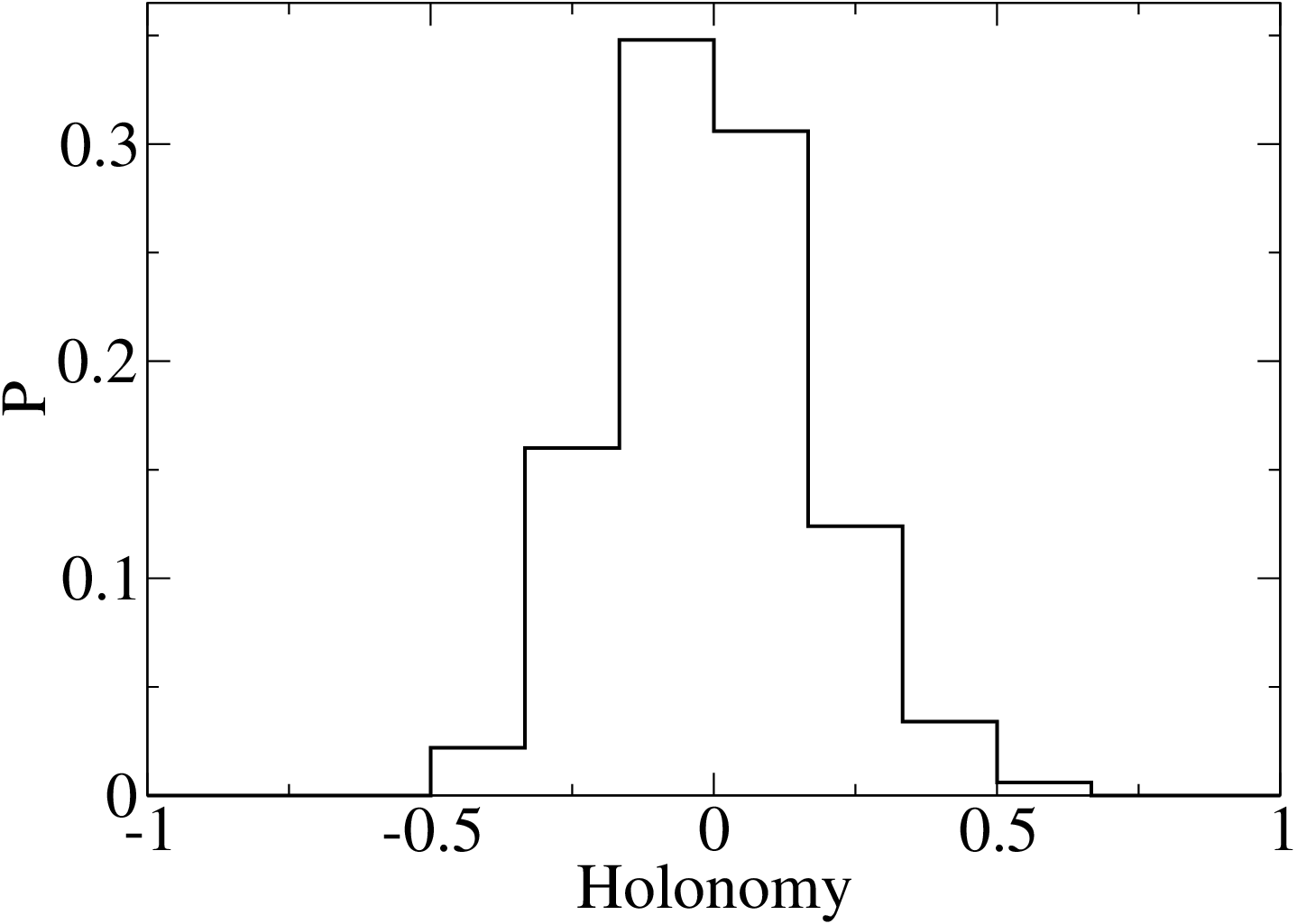}\\
\hspace{2.7cm} (a) \hspace{5.3cm} (b) \hspace{5.3cm} (c) \hspace{2.0cm}
\caption{For confining configurations on a $20^3 \times 6$ lattice at 
$\beta=2.3$ after $4D$ smearing, we show in:
(a) the distribution of the Polyakov loop in lattice points where 
time-like Abelian monopole currents are present after 100 smearing steps 
(solid histogram for the Polyakov gauge, dashed histogram for MAG) compared 
with the overall distribution of the Polyakov loop (dotted histogram) which 
is well described by the Haar measure (solid curve). All distributions are 
normalized.
(b) the distributions of the Polyakov loop in points where time-like 
Abelian monopole currents are present in Polyakov gauge comparing 100 and 50 
smearing steps (solid and dashed histograms correspondingly).
(c) the distribution of ``asymptotic holonomy'' (as defined in the text) 
after 100 smearing steps.}
\label{fig:plmoncorr}
\end{figure*}

Our first observation was the correlation between values of the Polyakov loop
and Abelian monopoles. The distribution of the Polyakov loop values in the 
points where Abelian monopole currents on timelike dual links
({\it i.e.} spacelike cubes occupied by Abelian magnetic charge) 
have been observed is shown in Fig. \ref{fig:plmoncorr}a after 
100 smearing steps. The solid and dashed histograms 
refer to monopoles in the Abelian projection corresponding to PG and MAG, 
respectively. The distribution is compared with the Polyakov loop distribution 
over all lattice points (dotted histogram) which is well described by the 
maximally random distribution 
derivable from the Haar measure, $P(L)=\sqrt{1-L^2}/(3\pi)$.

In Fig. \ref{fig:plmoncorr}b we compare the above mentioned distributions for 
lattice points with monopole currents (in the case of PG) obtained with 
respect to the number (100 and 50) of smearing steps. 
In order to facilitate a comparison of the clusters to analytical solutions 
with nontrivial holonomy, we define a so-called ``asymptotic holonomy'' for each 
lattice configuration. We consider the points on the lattice with the absolute 
value of topological charge density less than the averaged absolute value of 
topological charge density~\footnote{This definition is similar in spirit
but not the same as the definition that we have used before}. 
We take the average of the Polyakov loop over 
this set of ``asymptotic'' points.  The resulting distribution of the 
``asymptotic holonomy'' $H$ for 500 smeared configurations 
(after 100 smearing steps) is shown in Fig. \ref{fig:plmoncorr}c.
It can be seen from the last figure that the ``asymptotic holonomy'' $H$ is 
peaked at zero (understood as maximally nontrivial holonomy) and is 
still rather narrowly distributed. We will restrict further  
analysis to a subset of 330 smeared configurations with $|H| < 1/6$, 
{\it i.e.} with maximally nontrivial holonomy.

\section{Cluster analysis of smeared configurations}
\label{sec:cluster}
For each of the $4D$ smeared configurations satisfying the cut $|H| < 1/6$
we have looked for clusters of topological charge. The topological density 
is assigned to the lattice sites according to the plaquette definition.
We take the points where the absolute value of the topological charge density 
exceeds some threshold value. This threshold has been varied between the 
average absolute value of the topological charge density and a value taken
$10$ times larger.
The link-connected points above the threshold value (below minus the threshold 
value) form what we call the positive (negative) clusters of topological 
charge. The precise threshold itself for each  smeared configuration was 
chosen (within the above range) in such a way as to have the maximal number 
of disconnected clusters 
of topological charge for this configuration.

In this way we obtained $4684$ clusters, {\it i.e.} 
on average approximately 14 clusters per configuration. From these clusters 
we have selected $3464$ ($3107$) clusters that contain time-like 
Abelian monopole currents in PG (in MAG) as possible signatures for KvB 
monopole constituents. 
The remaining $1220$ ($1577$) clusters were free of time-like monopole currents.
Although all clusters together occupy on average only $3.5\%$ of the 4-d volume 
they contain $39.5\%$ of the time-like Abelian monopole currents 
detected on the lattice. 

Thus, time-like Abelian magnetic currents are about $18$ times more dense 
inside clusters of topological charge than outside. In order to select 
clusters containing either a single (more or less) static Abelian 
(anti)monopole, or monopole charges cancelling each other inside a 
topological cluster we determine an average monopole charge 
$~\overline{m}_{cluster}~$ for clusters containing timelike Abelian monopole 
currents. It is defined as the difference between the number of dual links 
with time-like currents going in the positive time direction 
(carrying positive magnetic charge) and the number of those with 
time-like currents going in the negative time direction 
(negative magnetic charge) divided by the total number of time-like 
monopole currents inside the cluster. 

In $60\%$ of clusters with Abelian monopoles we observed only equal-sign
monopole currents ({\it i.e.} all time-like Abelian monopole currents going
in the same direction) resulting in $~\overline{m}_{cluster} = \pm 1$. 
On the other hand in approximately $8\%$ of clusters the numbers 
of positive and negative time-like Abelian monopole currents were equal to 
each other, {\it i.e.} $~\overline{m}_{cluster} = 0$. 
Further, from the first group of clusters with 
$~\overline{m}_{cluster} = \pm 1~$ we have selected those 
with a number of time-like Abelian monopole currents larger than or equal 
to $N_{\tau}=6$ ($N_{\tau}$ is the number of time slices in the lattice), 
such that the Abelian monopole loop can close by periodicity in the time 
direction. Let us call them conditionally ``static'' clusters. In this way, 
finally, we have identified 547 (359) clusters with ``conditionally static'' 
monopoles in PG (MAG). Tentatively we labeled them as static dyons 
(eventually being part of a KvB caloron). 
The other 268 (638) clusters with a monopole-antimonopole pair are tentatively 
considered as undissociated KvB calorons. 

\begin{figure*}
\vspace{0.5 cm}
\centering
\includegraphics[width=.31\textwidth,height=.31\textwidth]{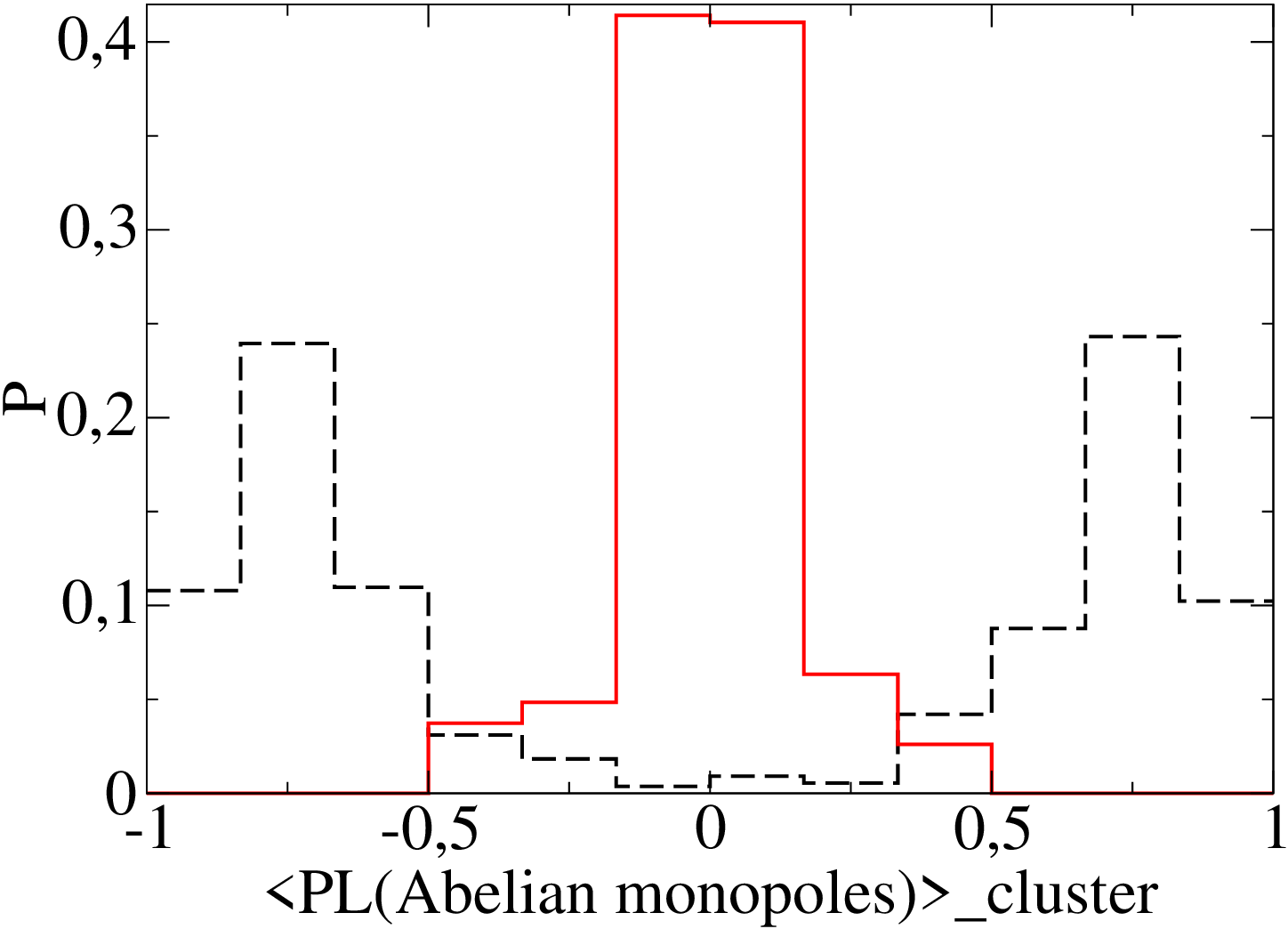}%
\hspace{.3 cm}\includegraphics[width=.31\textwidth,height=.31\textwidth]{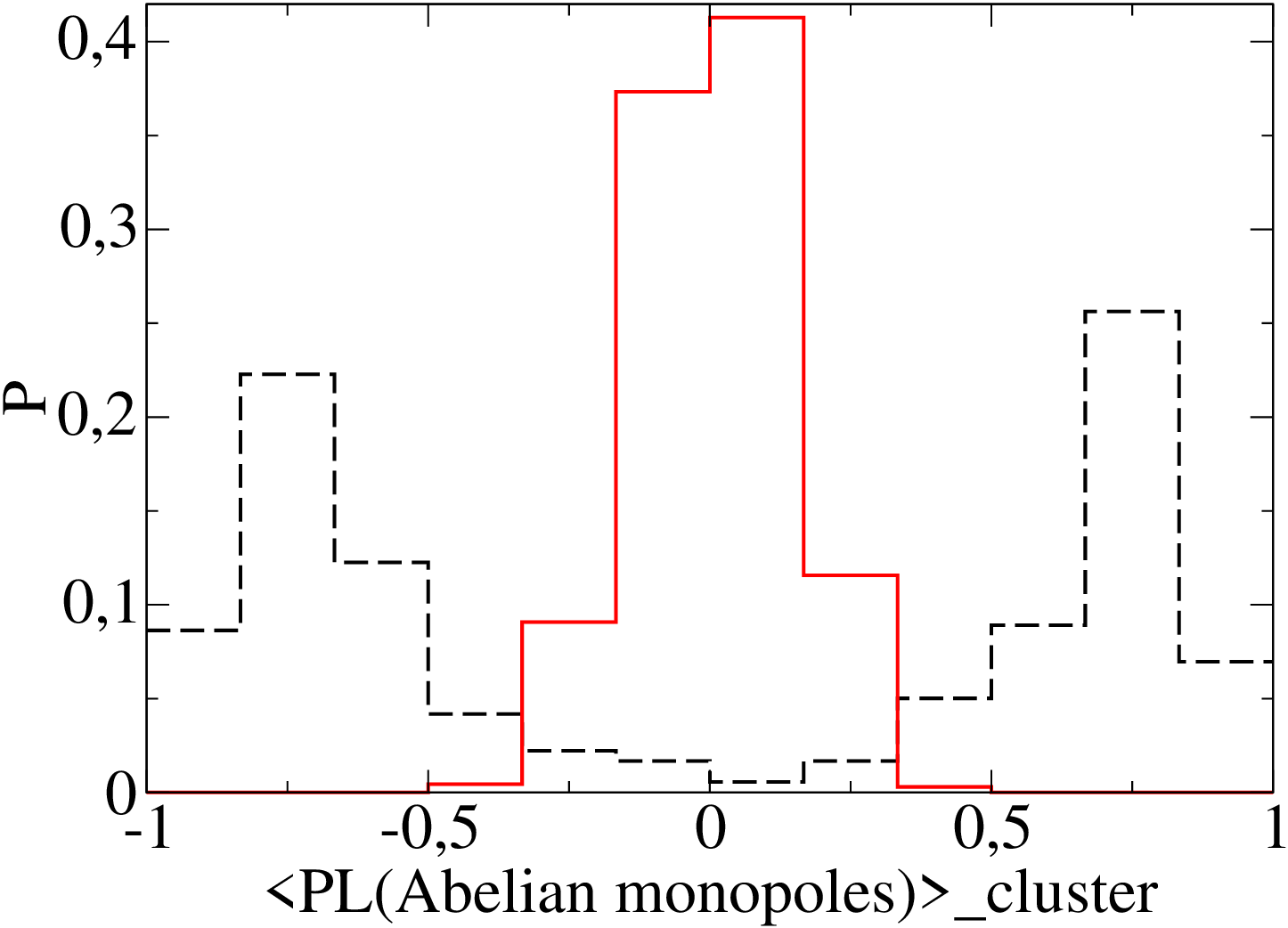}%
\hspace{.3 cm} \includegraphics[width=.31\textwidth,height=.31\textwidth]{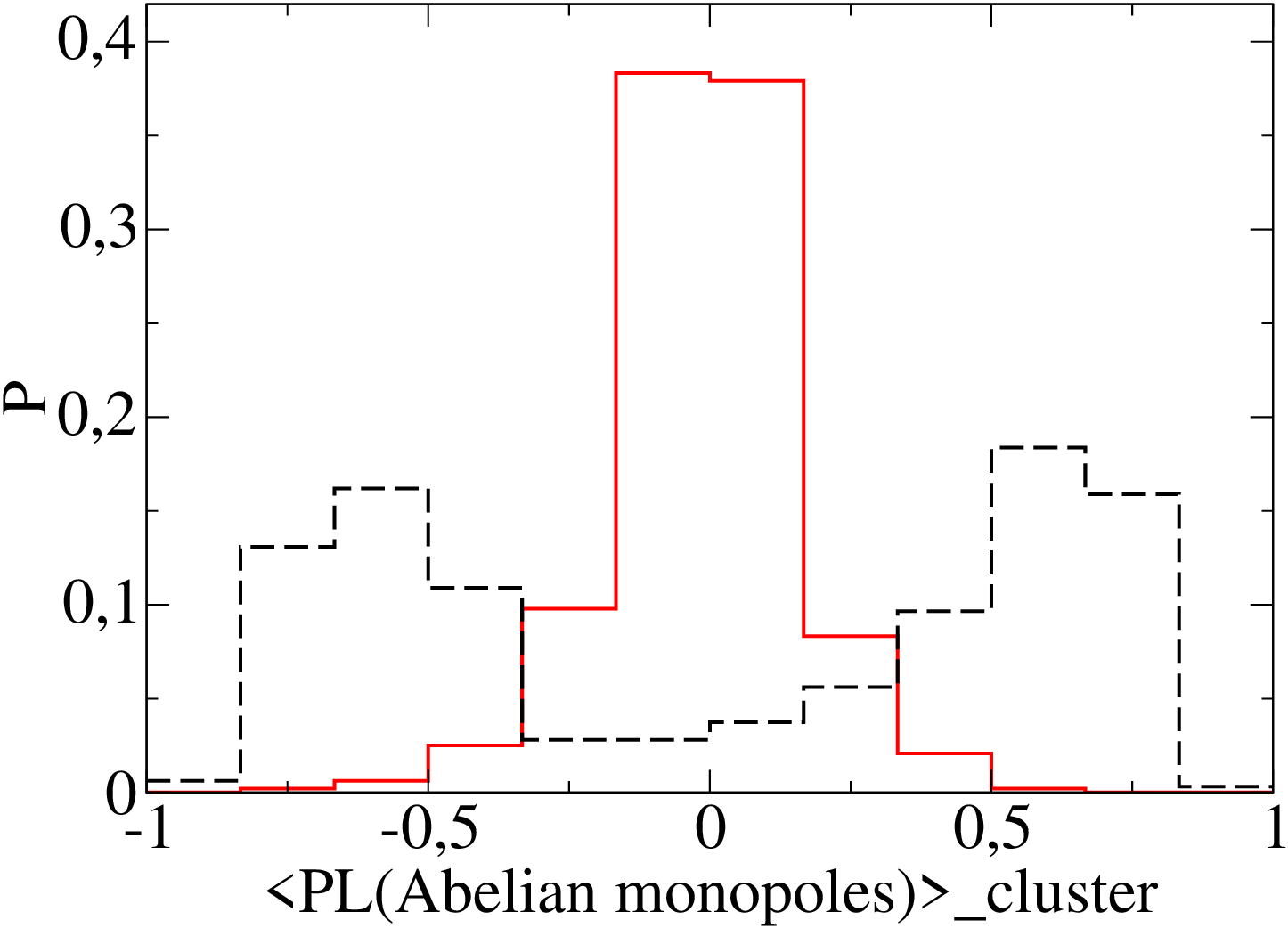}\\
\hspace{2.7cm} (a) \hspace{5.3cm} (b) \hspace{5.3cm} (c) \hspace{2.0cm}
\caption{(a) The distribution of the Polyakov loop averaged over the monopole 
skeleton of topological clusters with static monopoles (dashed histogram) and 
for clusters with monopole-antimonopole pairs (solid histogram)
(both according to the Polyakov gauge after 100 smearing steps). 
(b) The same for the maximally Abelian gauge and 100 smearing steps.
(c) The same for the Polyakov gauge after only 50 smearing steps.
}
\label{fig:plcldist}
\end{figure*}

In order to seek further support for this interpretation  
we have averaged the Polyakov loop inside the selected clusters  
over all points where time-like Abelian monopole currents of either
sign are observed. We call this quantity the averaged Polyakov loop of 
the monopole ``skeleton'' of the given cluster,
$\langle PL(\mathrm{Abelian~monopoles})\rangle_{cluster}$. 
From KvB calorons we expect this average for ``static monopole clusters'' 
($~\overline{m}_{cluster} = \pm 1~$) to be close to $~\pm 1~$, 
whereas for ``monopole-antimonopole pair clusters''
($~\overline{m}_{cluster} = 0~$) it should be close to $~0~$, the latter
because of the mentioned above dipole structure for the spatial Polyakov loop 
distribution inside the undissociated caloron.  
Indeed, the measured distributions of the Polyakov loop averaged over 
the monopole skeletons of the selected clusters are shown in 
Figs. \ref{fig:plcldist} are peaking around $\pm 0.75$ 
for clusters classified as static dyons and around zero
for clusters tentatively identified as undissociated KvB calorons. 
With fewer smearing steps the histogram for static dyons becomes less 
pronounced. 

Next we would like to get some information also about the topological charge of 
objects tentatively identified as static dyons and undissociated KvB calorons.
Since the clusters have been defined by means of a threshold for the density,
some (uncertain) part of the cluster charge is residing in the tail of the 
density and has to be appended to the charge integral over the cluster.
First we need some (model dependent) estimates for the actual size of the 
clusters of both kinds before we are able to define the total cluster charge 
by including the (observed) tail of the charge distribution as well.
These estimates are different for the two types of clusters.

For a static dyon we know from the analytic KvB caloron solution that its 
size depends on the holonomy according to $~\bar{r}={b}/{4\pi\omega}~$
($\omega~$ is the holonomy parameter, $H=\cos (2\pi\omega)$ and $b$ is
the inverse temperature, the period in time direction).
The topological charge density in the center, $q_{max}$, scales with the 
size in the following way:
\begin{equation}
q_{max} = \frac{1}{24\pi^2\bar {r}^4}~~.
\label{eq:dyondens}
\end{equation}
(see the Appendix for some details).
Fitting clusters classified as static dyons to this equation
we can infer the cluster size from the 
observed maximum of the topological charge density inside the cluster.
Then we sum the topological charge of all points that have a {\it spatial} 
distance from the point of maximum less than some radius $~R~$ related to 
$~\bar{r}$.
This distance $~R~$ should not be too large in order to 
avoid double counting of topological charge density 
(by assigning points to more than one cluster) and  
not too small (in order not to underestimate the topological charge of 
the cluster considered). We use $R=3\bar{r}$ which would give for an 
isolated cluster (with an ideal, exponential profile of the topological 
charge density) the total charge within $7\%$ accuracy (there is no need 
to correct for the tail). In this way we 
assign a topological charge to all clusters classified as static dyon 
clusters.

An undissociated KvB caloron has a topological charge profile like that of
an isolated ordinary instanton solution. 
For them the maximum of the topological charge 
density is related to the instanton size $~\rho~$ as follows
(see Appendix)
\begin{equation}
q_{max} = \frac{6}{\pi^2\rho^4}~~.
\label{eq:instdens}
\end{equation}
Assuming that the clusters classified as undissociated calorons have this 
charge profile we can obtain the instanton 
size $~\rho~$ from the measured $~q_{max}~$ of the cluster. 
Then we sum the topological charge over all points that have a 
{\it 4-dimensional} distance less than $1.5\rho$ from the maximum position.
The result needs to be  multiplied by a correction factor $~1.29~$ as for 
the exact instanton solution (see Appendix).
In this way we  define an estimated topological charge also for clusters 
identified as undissociated calorons.

At this point we wish to explain why 100 smearing steps are more suitable
than 50 for the detection of KvB dyons. Indeed, we will show 
later that the signal becomes more clear with more smearing steps.
For an isolated dyon the topological density in the maximum is equal to
$|q|_{max} = \pi^2/(24 \times 6^4)~~$ as it can be seen from 
Eq. (\ref{eq:dyondens}) 
with $\omega=1/4$ (the case of maximally nontrivial holonomy) and $b = 6$. 
Requiring the maximum of topological density in a cluster to exceed the 
threshold value ({\it i.e.} the averaged modulus of the topological density 
over the configuration), $|q|_{max} > \overline{|q|}$, we get an
upper limit for $\overline{|q|}$ of a smeared configuration in order 
for an isolated dyon with maximally nontrivial holonomy to be recognized 
as a cluster by our cluster finding algorithm.
This would give an action value $|q|_{max} \times 6 \times 20^3 \approx 15$ 
as an upper limit
for the smeared action (given in instanton units). This value is close to the
action of smeared configurations after 100 smearing steps as mentioned 
before.

\begin{figure*}
\vspace{0.5cm}
\centering
\includegraphics[width=.31\textwidth,height=.31\textwidth]{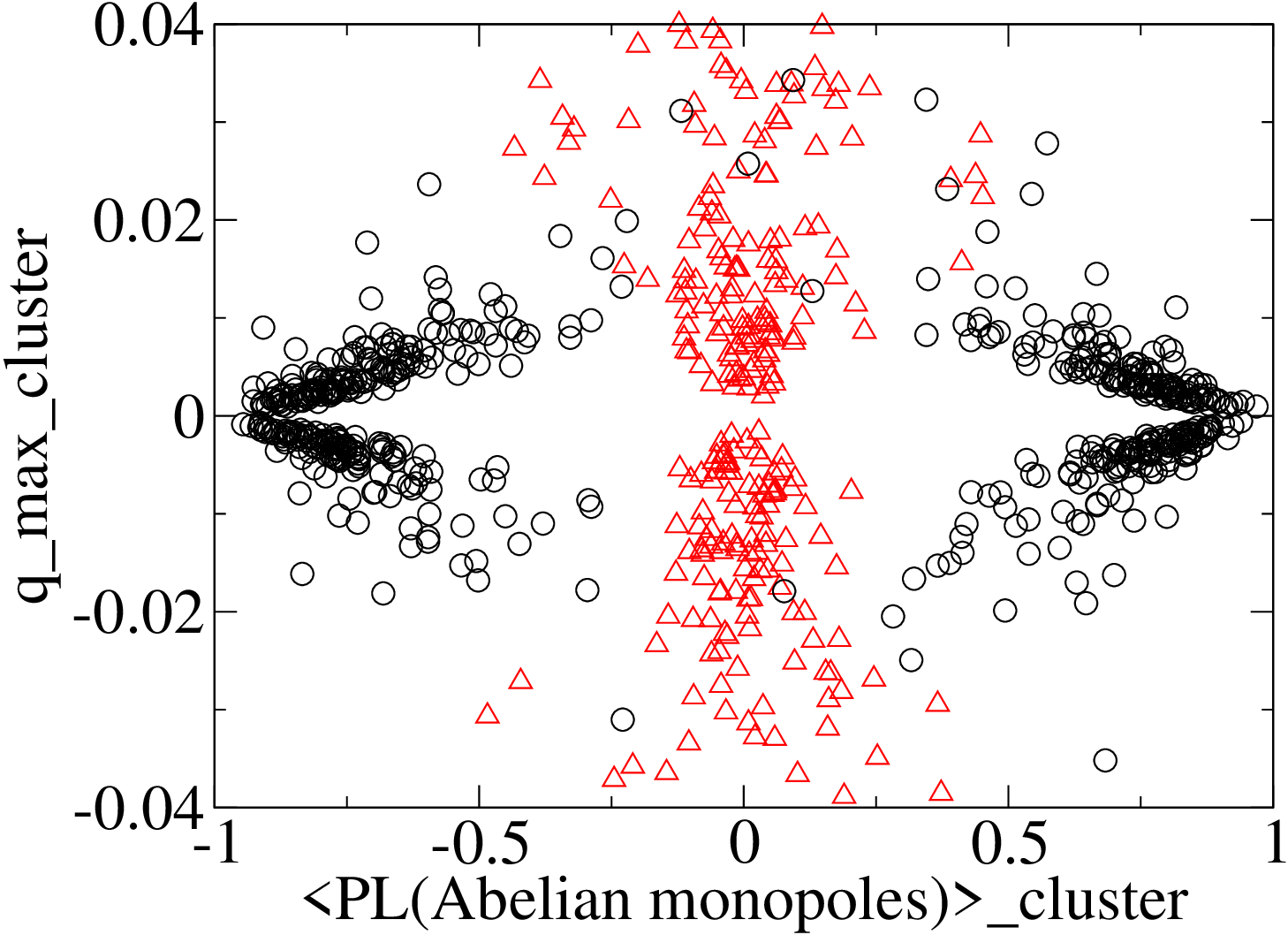}%
\hspace{.3
cm}\includegraphics[width=.31\textwidth,height=.31\textwidth]{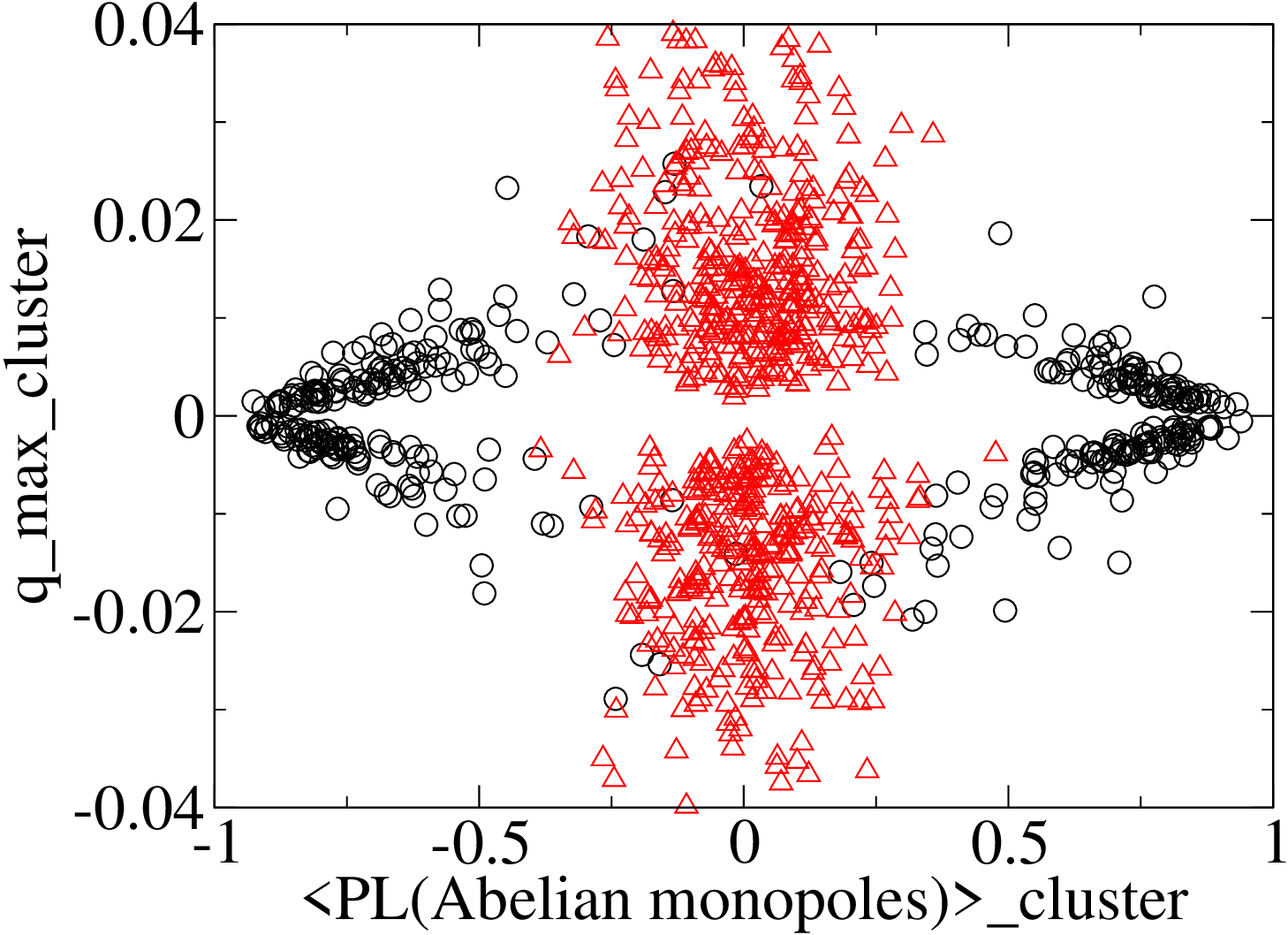}%
\hspace{.3
cm}\includegraphics[width=.31\textwidth,height=.31\textwidth]{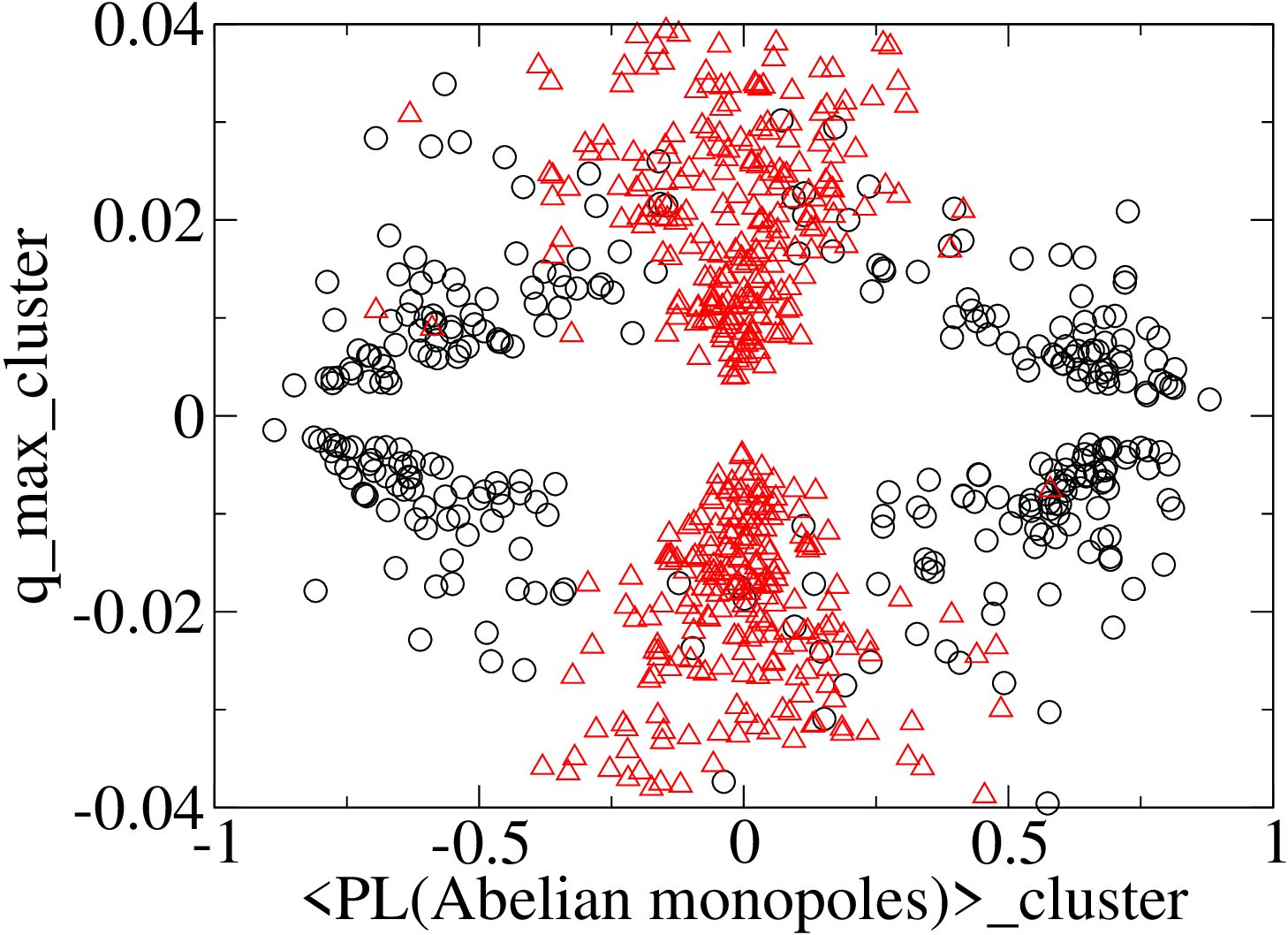}\\
\hspace{2.7cm} (a) \hspace{5.3cm} (b) \hspace{5.3cm} (c) \hspace{2.0cm}
\caption{(a) The scatter plot shows the concentration of topological clusters
after 100 smearing steps in the 
${(q_{max})}_{cluster}$-$~\langle PL(\mathrm{Abelian~monopoles})\rangle_{cluster}$ 
plane, depending on the monopole content in the Polyakov gauge: 
clusters with static monopoles
(circles) and clusters with monopole-antimonopole pairs (triangles) are forming
different clusters in this plot.
(b) The same for the maximally Abelian gauge.
(c) The same for the Polyakov gauge after only 50 smearing steps.
}
\vspace{-0.3cm}
\label{fig:pldqcorr}
\end{figure*}

As for the PG, we can present the 
two sorts of topological clusters ($547$ topological clusters seen after 100
smearing steps classified as static (anti)monopoles 
and $268$ clusters interpreted as monopole-antimonopole pairs) 
in a scatter plot with respect to the maximal value $~q_{max}~$ on one hand 
and the averaged Polyakov loop of the monopole skeleton 
($\langle PL(\mathrm{Abelian~monopoles})\rangle_{cluster}$)
on the other in Fig. \ref{fig:pldqcorr}a. The same for MAG is shown in 
Fig. \ref{fig:pldqcorr}b. The dependence on the number of smearing steps for the 
PG case can be concluded from Fig. \ref{fig:pldqcorr}c which refers to 50 smearing 
steps. The average size of single (anti)monopoles after 100 smearing steps is 
$~\bar{r} \approx (1 \ldots 1.5)~a~$, 
whereas the most probable size of undissociated 
calorons is $~\rho \approx 3~a$ corresponding to a distance between 
constituents $d \approx 4.5~a$. The lattice spacing $~a~$ 
can be related to the deconfining temperature according to 
$~\beta=2.30 \approx \beta_c(N_{\tau}=4)=2.29~$, resulting in
$~a \approx 1/(4 T_c)$ . Less smearing results (on average) in a
smaller size of both types of clusters (causing problems due to the finite
resolution of the lattice).

\begin{figure*}
\vspace{0.8cm}
\centering
\includegraphics[width=.31\textwidth,height=.31\textwidth]{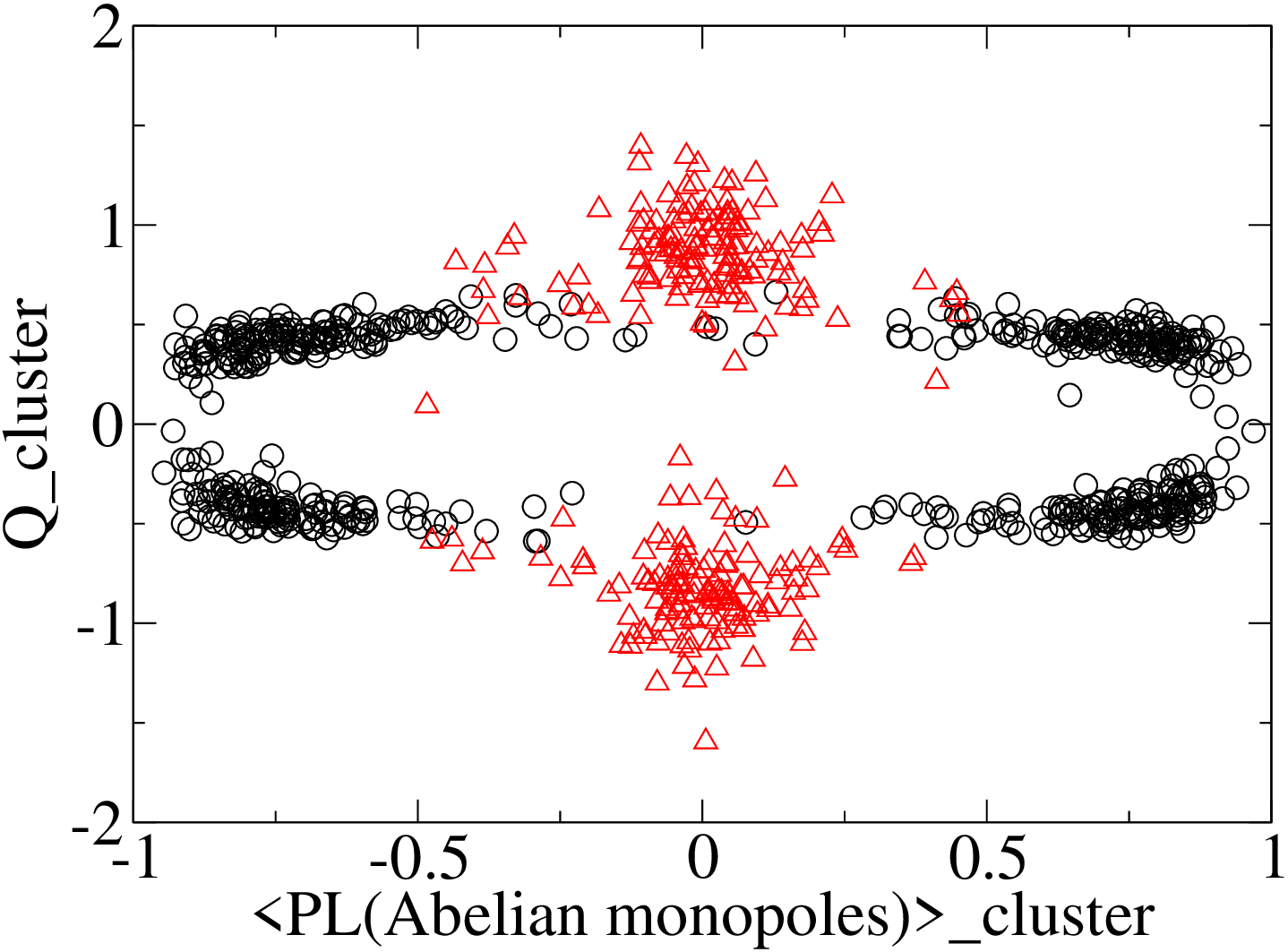}%
\hspace{.3 cm}\includegraphics[width=.31\textwidth,height=.31\textwidth]{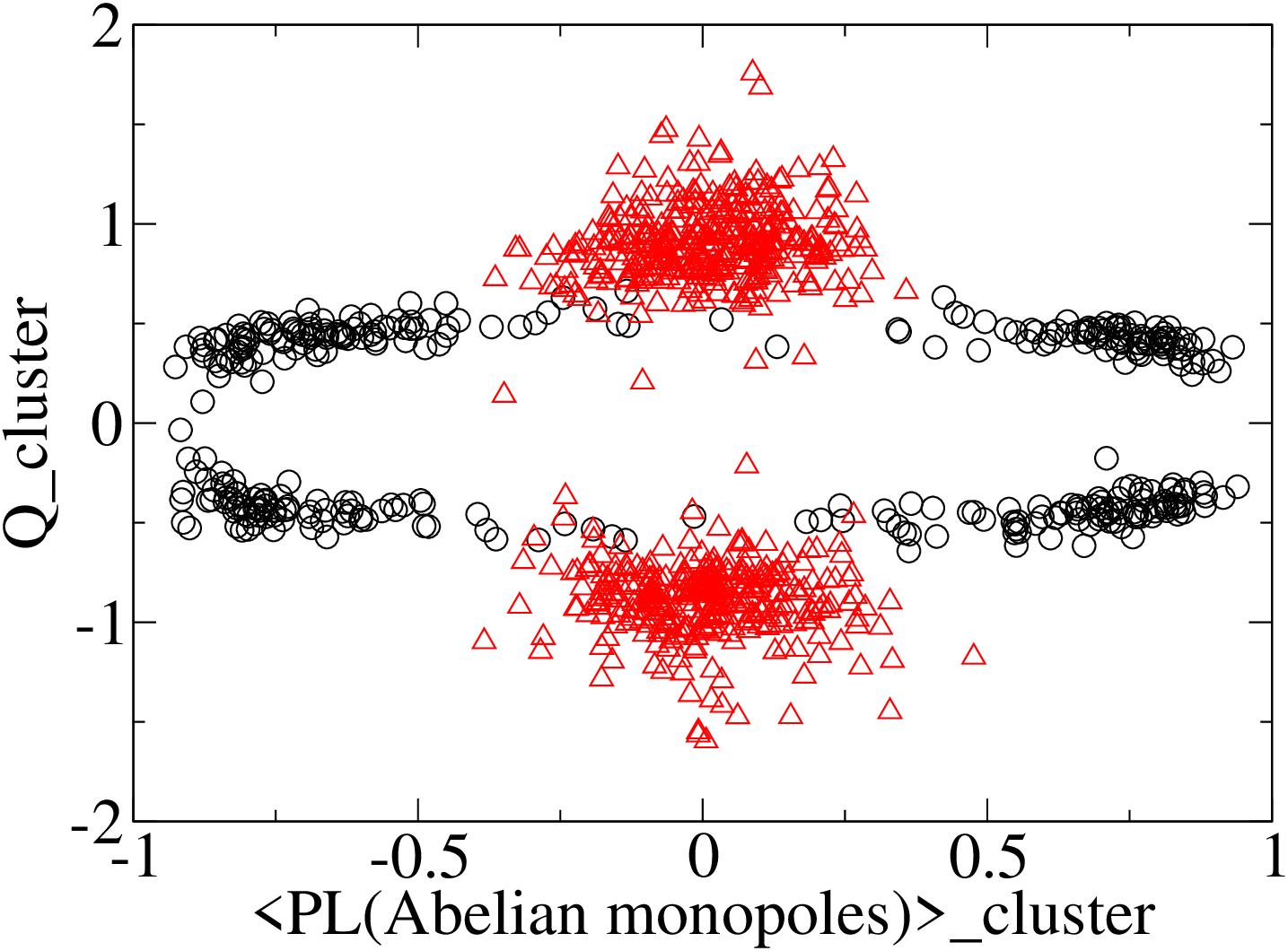}%
\hspace{.3 cm}\includegraphics[width=.31\textwidth,height=.31\textwidth]{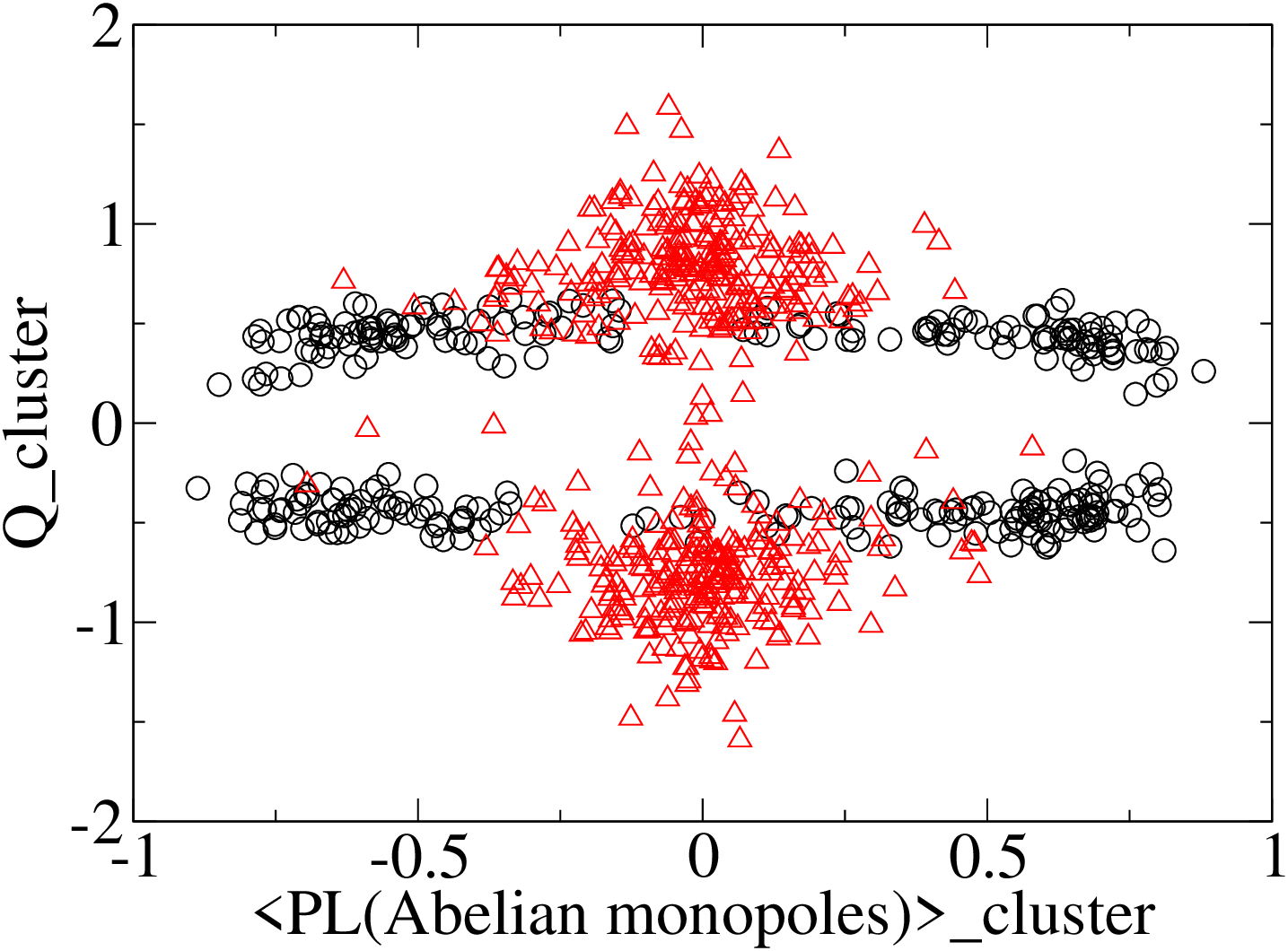}\\
\hspace{2.7cm} (a) \hspace{5.3cm} (b) \hspace{5.3cm} (c) \hspace{2.0cm}
\caption{Scatter plots as in Figs. \ref{fig:pldqcorr} (a), (b) and (c), respectively, 
but for the 
$Q_{cluster}$-$~\langle PL(\mathrm{Abelian~monopoles})\rangle_{cluster}$ plane.
}
\label{fig:plqcorr}
\end{figure*}

The corresponding scatter plots with the estimated topological charge of each 
cluster $~Q_{cluster}~$ on one hand and the averaged Polyakov line of the 
monopole skeleton ($\langle PL(\mathrm{Abelian~monopoles})\rangle_{cluster}$)
on the other are presented in Fig. \ref{fig:plqcorr}.
As it can be seen from this Figure the two sorts of topological 
clusters are clustering on the scatter plot either near the points 
$~\langle PL(\mathrm{Abelian~monopoles})\rangle_{cluster} =\pm 1, 
~Q_{cluster} = \pm 1/2~$ (dissociated) or 
$~\langle PL(\mathrm{Abelian~monopoles})\rangle_{cluster} = 0,
~Q_{cluster} = \pm 1~$ (undissociated). 

\begin{figure*}
\vspace{0.5 cm}
\centering
\includegraphics[width=.31\textwidth,height=.31\textwidth]{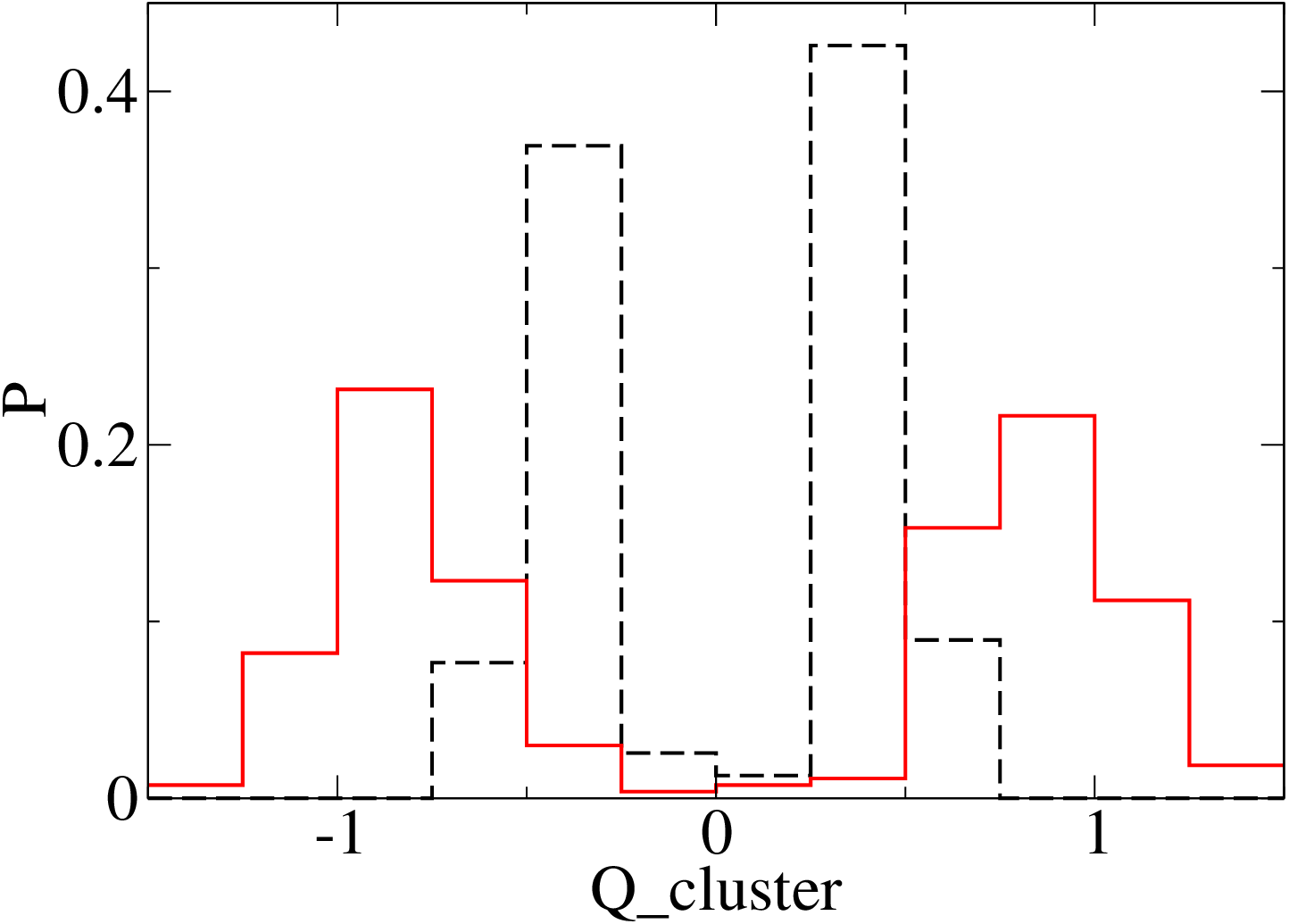}%
\hspace{.3 cm}\includegraphics[width=.31\textwidth,height=.31\textwidth]{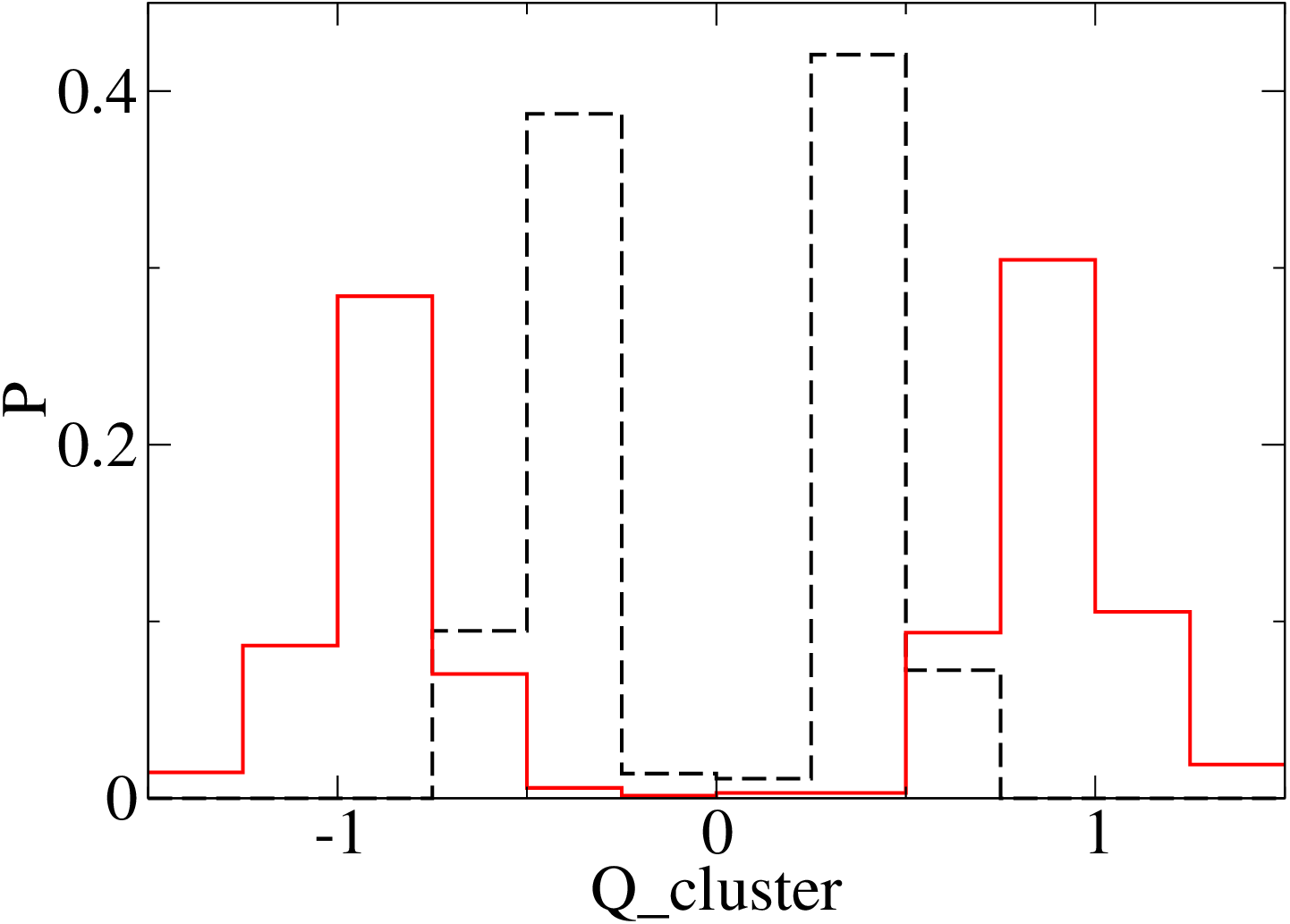}%
\hspace{.3 cm}\includegraphics[width=.31\textwidth,height=.31\textwidth]{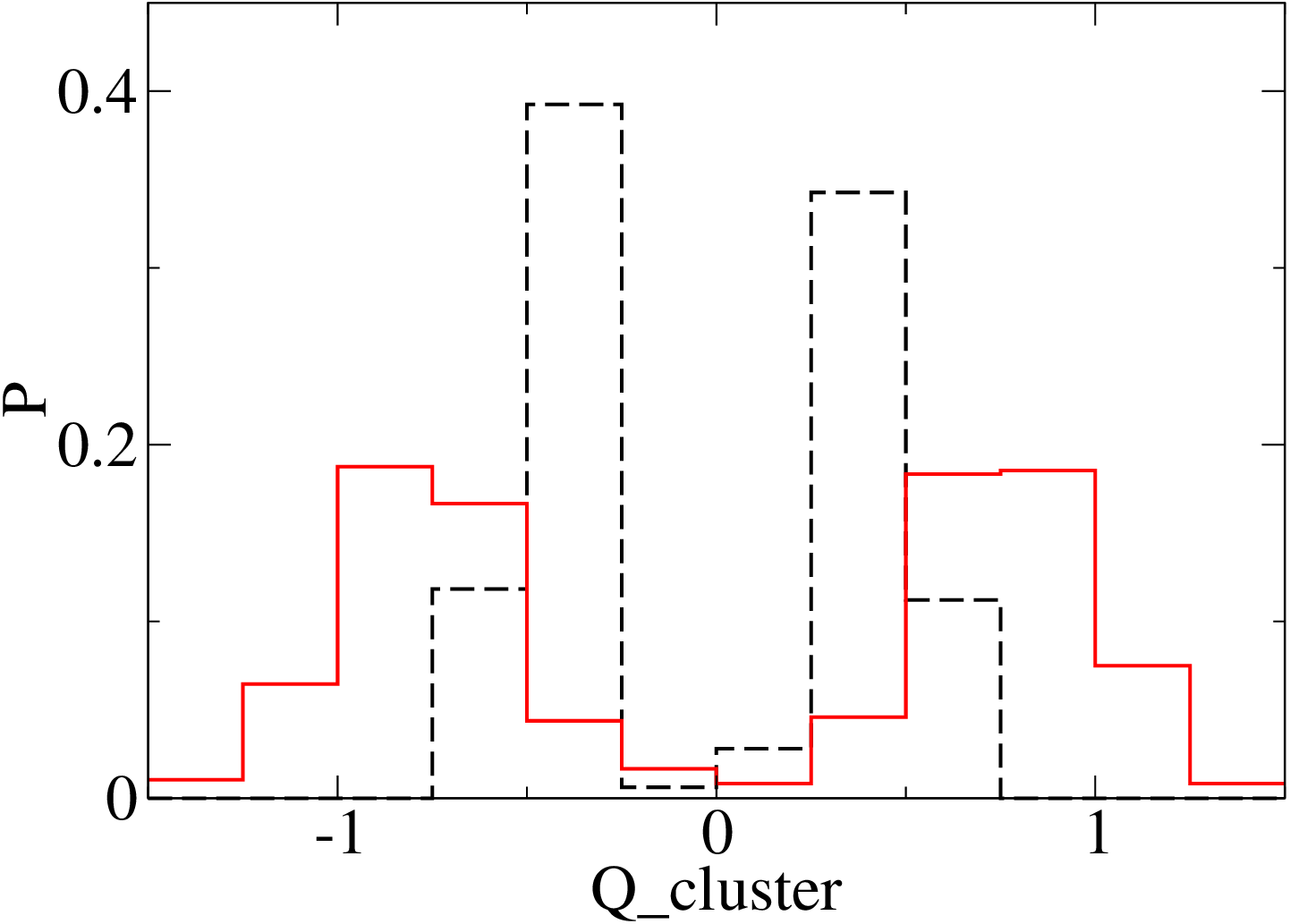}\\
\hspace{2.7cm} (a) \hspace{5.3cm} (b) \hspace{5.3cm} (c) \hspace{2.0cm}
\caption{(a) The distribution of the topological charge per cluster, 
$Q_{cluster}$, of topological clusters after 100 smearing steps with static 
monopoles (dashed histogram) and for clusters with monopole-antimonopole pairs 
(solid histogram) in the Polyakov gauge.
(b) The same for the maximally Abelian gauge.
(c) The same for the Polyakov gauge after only 50 smearing steps.
}
\label{fig:qcldist}
\end{figure*}

The existence and interpretation of these two sorts of topological clusters 
can also be concluded from the corresponding reduced distributions shown on
Fig. \ref{fig:plcldist}a and Fig. \ref{fig:qcldist}a focussing on 
$\langle PL(\mathrm{Abelian~monopoles})\rangle_{cluster}$ and $Q_{cluster}$, 
respectively, all for PG.
Figs. \ref{fig:plqcorr}b, \ref{fig:plcldist}b and \ref{fig:qcldist}b 
refer to MAG, while
Figs. \ref{fig:plqcorr}c, \ref{fig:plcldist}c and \ref{fig:qcldist}c 
describe PG after 
only 50 smearing steps, in order to demonstrate the effect of smearing.

\begin{figure*}
\vspace{0.5 cm}
\centering
\includegraphics[width=.31\textwidth,height=.31\textwidth]{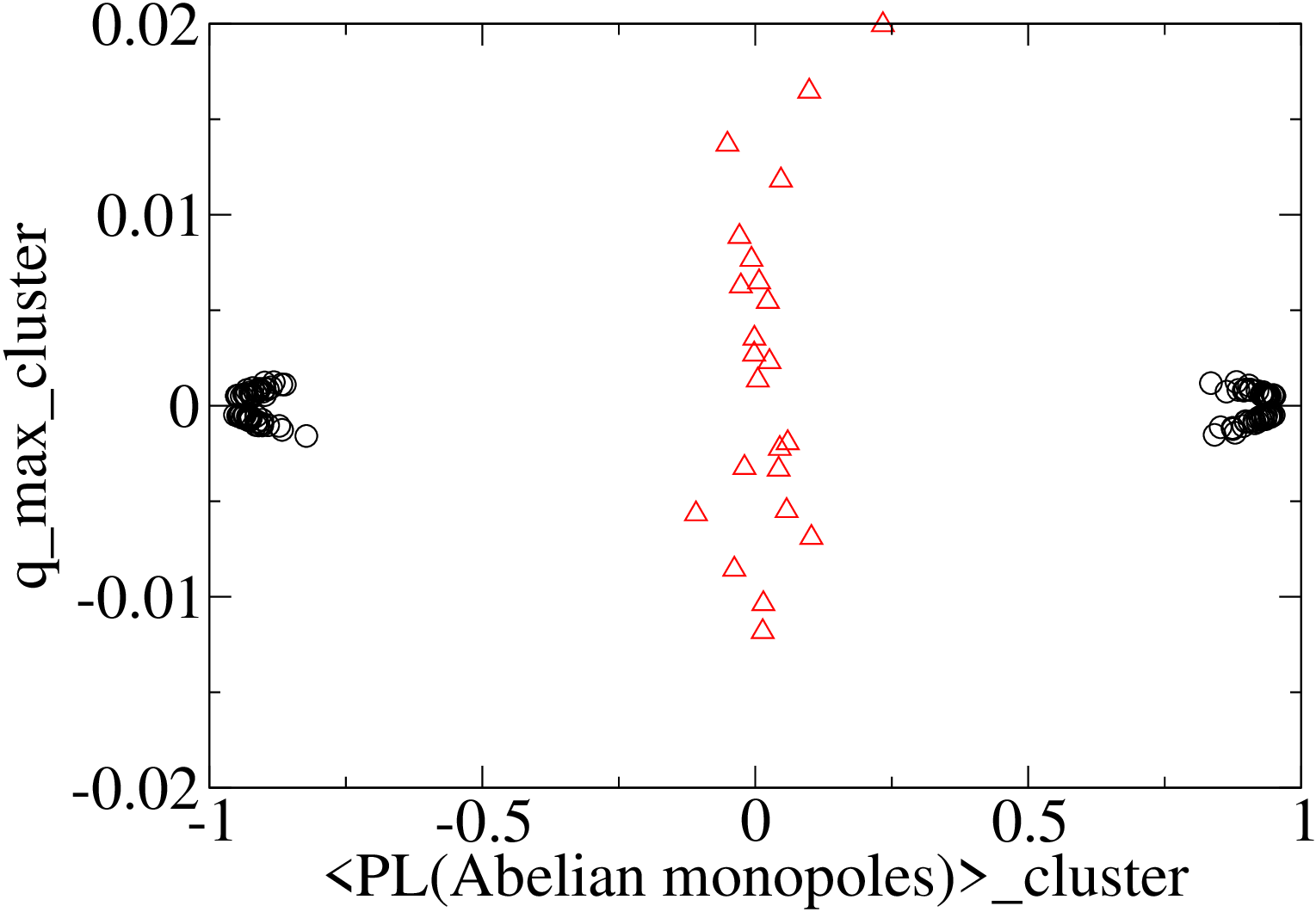}%
\hspace{.5cm}\includegraphics[width=.31\textwidth,height=.31\textwidth]{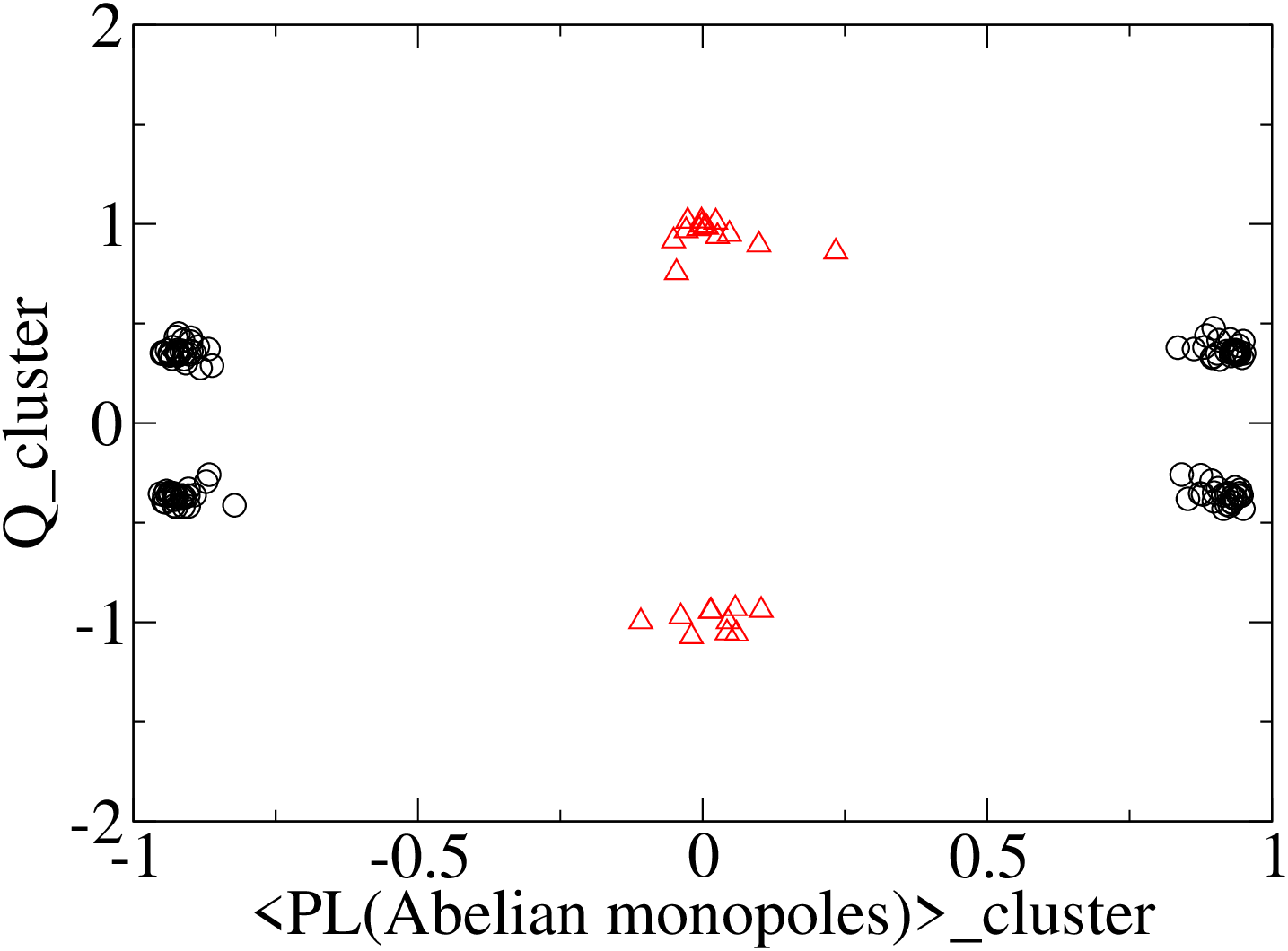}\\
\hspace{0.6cm} (a) \hspace{5.2cm} (b) 
\caption{Scatter plots as in Figs. \ref{fig:pldqcorr} and \ref{fig:plqcorr}~~ but 
for $O(100)$ of artificially and randomly generated single caloron configurations. 
The monopole content is defined in the maximally Abelian gauge. }
\label{fig:artif_KvB}
\end{figure*}

That is what we expected for isolated dyons from KvB solutions in a background
of maximally nontrivial holonomy and for undissociated KvB calorons in the same
background. In order to check our picture we generated artificial topological 
clusters by discretizing analytical single KvB caloron solutions with maximally 
non-trivial holonomy also on a $20^3 \times 6$ lattice. 
They were subjected to a few improved cooling steps in order to 
adapt them to the 3-dimensional periodicity of the lattice.
The distance of the constituents was randomly varied 
from zero to the maximal possible value of $10 \cdot \sqrt{3}$,
in order to create both dissociated and non-dissociated 
calorons~\footnote{The exact separation between these cases
does not matter for this purpose.}
We investigated these artificial clusters with the same instruments as 
described above. The results of this model calculation using MAG for detecting
the Abelian monopoles are visualized in Figs. \ref{fig:artif_KvB} and agree nicely 
with our findings before. 

Thus, we can conclude to have found a clear signal (in smeared $SU(2)$ lattice 
configurations in the confining phase) of topological objects 
falling under the classification offered by extreme cases of the KvB solutions. 
One may wonder why only $~547+268=815~$ clusters (in PG) from $~4684~$ clusters in 
total are distinguished by their discernible monopole and topological content. 
It should be taken into account that the above signal could be clearly seen 
only for well isolated clusters, and we have selected clusters according to
extreme cases of well separated constituents and instanton-like calorons.
Generically, the clusters are just mutually disconnected,
and there is no such clear relation between the maximal action density and 
the size of the cluster.

\section{Conclusion}
\label{sec:conclusion}
Investigating equilibrium lattice fields obtained at finite temperature 
in $SU(2)$ gluodynamics we have demonstrated that among the topological 
objects (observed in the confined phase after suitable smearing) 
there are both static dyons and nonstatic calorons.
Static dyons are correlated with static Abelian monopoles obtained from
Abelian projection in Polyakov gauge or maximally Abelian gauge. 
Nonstatic calorons are correlated with nonstatic loops of Abelian 
monopole-antimonopole pairs. The behaviour of the local Polyakov loop inside 
these objects and the (model-dependent) estimates of their topological charges 
favour the interpretation as KvB calorons with nontrivial holonomy, 
or as constituent dyons into which such KvB calorons can dissociate.

\vspace{2.0cm}

\section*{Appendix}
The purpose of this appendix is to explain the method used to
estimate the topological charge of objects first detected as clusters. The 
estimate depends on whether they have been classified as static dyons or 
undissociated KvB calorons.

Undissociated KvB calorons have an action profile like
an ordinary instanton. For the latter the action density is
equal to
\begin{equation}
s(x) = \frac{48 \rho^4}{(x^2 + \rho^2)^4}~.
\label{eq:actdensinst}
\end{equation} 
The value $~s(0)~$ normalized to the total instanton action of
$~S_{inst}=8\pi^2~$ is presented in the text (see Eq. (\ref{eq:instdens})).
If one integrates the above action density over the 4-dimensional volume
bounded by a sphere around the center with the radius $~1.5 \rho~$ the 
result is equal to $~S_{inst}/1.29$. Therefore, in order to get the total 
action (topological charge) of an undissociated caloron we have to correct 
the above restricted integral (sum over lattice points) by the factor $~1.29$.

The other extreme case of an isolated dyon is more involved~\cite{0404210}. 
The action density for the general caloron solution with nontrivial holonomy 
is given by the following formulae \cite{KvB-1}
\begin{eqnarray} 
\label{eq:solutionact}
 s(x)& = &-\frac{1}{2}\partial^2_\mu\partial^2_\nu \log \psi(x) \,\nonumber\\ 
\psi(x) &=& \cosh(4\pi r\omega)\cosh(4\pi s \bar{\omega}) \\
 & + & \frac{(r^2\!+\!s^2\!+\!\pi^2\rho^4)}{2rs}
   \sinh(4\pi r \omega) \sinh(4\pi s\bar{\omega})  \nonumber \\
 & + & \pi\rho^2 \left( s^{-1}\sinh(4\pi s\bar{\omega})\cosh(4\pi r\omega) \right. \nonumber\\
 & &  + \left. r^{-1}\sinh(4\pi r\omega)\cosh(4\pi s\bar{\omega})\right) 
 \nonumber\\
 & - & \cos(2\pi t)~\nonumber ,
\end{eqnarray}
where the period in time direction is set equal to $b=1$
(in other words, all distances are measured in $b$).
The holonomy parameters $\omega$ and $\bar{\omega}$ are related to each other 
$\bar{\omega} = 1/2 - \omega, ~~0 \le \omega \le 1/2$. The distances
$~r=|\vec{x}-\vec{x}_1|~$ and $~s=|\vec{x}-\vec{x}_2|~$
are the 3-dimensional distances from the locations of the two centers of the 
caloron solution. The distance between the centers 
$~d \equiv |\vec{x}_1-\vec{x}_2|~$ is connected with the scale size and the
width of the time periodicity strip $~b~$ through
\begin{equation} 
\label{eq:drho}
\pi\rho^2/b = d \,.
\end{equation} 
Now if we remove the second center (the dyon at $~\vec{x}_2~$) to infinity
$~s \approx d \rightarrow \infty~$ we can find the potential $~\psi~$ for
an isolated dyon, which obviously is static
\begin{equation}
\label{eq:psi}
\psi(x) = A\frac{\sinh(4\pi r \omega)}{r}~,
\end{equation}
where $~A~$ is some (infinite) constant not important for the calculation
of the action density. Invoking the inverse temperature $~b~$ again,
$~\bar{r}={b}/{4\pi\omega}~$
can be called the size of the dyon.
In order to obtain the action density at the position $~\vec{x}_1~$ we
expand $~\log \psi(x)~$ up to the fourth power in $~r~$
\begin{equation}
\label{eq:log}
\log \psi(x)= A^\prime +\frac{1}{6}\frac{r^2}{\bar{r}^2}-\frac{1}{180}
\frac{r^4}{\bar{r}^4} + ...
\end{equation}
and calculate the fourth derivative 
\begin{equation}
\label{eq:fourthderiv}
s(\vec{x}_1)= \frac{1}{3~\bar{r}^4}~.
\end{equation}
Eq. (\ref{eq:dyondens}) in the text is obtained by normalization
of $~s(\vec{x}_1)~$ to the total instanton action $~S_{inst}=8\pi^2$.

\section*{Acknowledgements}

This work was partly supported by RFBR grants 02-02-17308, 03-02-19491
and 04-02-16079, DFG grant 436 RUS 113/739/0 and RFBR-DFG grant 03-02-04016 and
by Federal Program of the Russian Ministry of Industry,
Science and Technology No 40.052.1.1.1112. 
Two of us (B.V.M. and A.I.V.) gratefully appreciate the support of 
Humboldt-University Berlin where this work was carried out to a large extent. 
E.-M. I. is supported by DFG (FOR 465 / Mu932/2). The authors acknowledge
constructive remarks by P. van Baal, F. Bruckmann and C. Gattringer.



\end{document}